# Effects of Initial State on Opinion Formation in Complex Social Networks with Noises

Yi Yu, Vu Xuan Nguyen, Gaoxi Xiao

*Abstract*— Opinion formation in complex social networks may exhibit complex system dynamics even when based on some simplest system evolution models. An interesting and important issue is the effects of the initial state on the final steady-state opinion distribution. We show that, while different initial opinion distributions certainly make differences to opinion evolution in social systems without noises, in systems with noises, given enough time, different initial states basically do not contribute to making any significant differences in the final steady state. Instead, it is the basal distribution of the preferred opinions that contributes to deciding the final state of the systems. Such an observation contradicts with a long-term belief on the roles of system initial state in opinion formation. We propose some brief discussions on the reasons supporting our statement, and the implications of such an observation in real-life applications.

## I. Introduction

OPINION formation is an important and interesting topic in studies on complex networks, especially complex social networks [1, 2]. It is of practical importance to predict how different opinions may evolve in the system and which opinions may dominate in a population after a temporal process of spreading through social interactions [3, 4]. Existing studies have revealed that, given enough time, local pair-wise interactions of individuals, in most cases, may eventually lead to the formation of a global equilibrium.

Quite a few models have been proposed to reveal system dynamics of opinion formation by introducing various rules of communication between individuals. Some of the most well-known ones include Voter model [5, 6], major rule model [7], Sznajd model [2], etc. Proposed by Deffuant *et al.* in 2000, the Deffuant model [8], as one of the most popular models, has been extensively studied. It is found that the tolerance bound (also termed as confidence bound, tolerance range, and uncertainty threshold etc. [9–11]), which reflects how much people may tolerate different opinions and make consensus, plays an important role in opinion formation processes.

In the classic Deffuant model, there is a continuous distribution of different opinions and a randomly chosen node may make consensus with a randomly selected neighbor holding *similar* opinion; their opinions hence come closer to each other or become the same. Individuals holding significantly different opinions, on the other hand, may not easily achieve an agreement. Link rewiring was later introduced into the Deffuant model [10], where two connected individuals holding significantly different opinions (terms as dissenters) may choose to cut the link in between (a system with link rewiring typically reflects a lower tolerance level compared to that without link rewiring) and connect with a similar opinion holder instead. It is shown that the existence of rewiring makes it harder for an adaptive network to reach global consensus; network may finally evolve into a few big opinion communities, each of which holding its community consensus [10].

Various factors other than consensus making between directly connected individuals, e.g., education, propaganda, community activities, etc. may also contribute to opinion changes. The combined effects of all these were introduced in [12, 13], termed as "*noises*". A simplest model of such noises is indeed to model the combined effects as random noises. Simple as the model is, it nevertheless helps reveal the nontrivial roles that such effects may impose on the system opinion evolution. It is shown that in the Deffuant model with a continuous opinion distribution and a fixed network topology, different speeds of random opinion change may drive the system either to an ordered state with a set of well-defined opinion groups, or a disordered state where the opinion distribution tends to be uniform. Another thread of research on random opinion changes, sometimes term such changes as *opinion mutation* (hereafter "opinion mutation" and "opinion noises" will be used interchangeably), confirms that such changes play a critical role in deciding the final steady state of system's long-term opinion evolution [14, 15]. In fact, it is the basal distribution of preferred opinion in opinion mutation that largely decides the final steady state of the system [13, 14].

When consensus making, link rewiring and opinion mutation are all taken into account in system evolution, some surprisingly complex system behaviors may emerge [16]. Most noticeably, it is shown that the complex dynamics may lead to different numbers of opinion communities at steady state with a given tolerance level between different opinion holders.

A small but nontrivial issue that has not been fully understood is the effects of initial state, i.e., initial opinion distribution, on the final steady of opinion distribution in complex social networks. Specifically, when there is no

Yi Yu is with Data Science Innovation Hub, Merck Sharp & Dohme, Singapore (email: olopgh@hotmail.com).
Vu Xuan Nguyen is with School of Electrical and Electronic Engineerting, Nanyang Technological University, Singapore (email nguyenxu001@e.ntu.edu.sg).

Gaoxi Xiao is with School of Electrical and Electronic Engineering, Nanyang Technological University, Singapore (corresponding author, phone: 65-67904552, email: egxxiao@ntu.ediu.sg)

mutation of opinion in the system, as that has been studied and concluded in [13], different initial states may lead to different final states. This can be easily understood: when there is no opinion mutation, different opinions may evolve into a number of isolated communities with no connections in between. When the community forming-up is fast enough, different initial states may make differences to the final states. For example, an initial state where majority of population holds opinions biased to right-hand side may end up with having a final state where the majority still holds opinions biased towards right-hand side.

When there exists mutation in opinion evolution, though our own observations and existing results both agree that the opinion mutation makes the effects of the initial state on the final state much less significant than those of the cases without mutation, we are not drawing exactly the same conclusion: in the existing results, it is believed that different initial states may still lead to different final states [13], while our observations, as later we shall report in detail in this paper, show that such differences virtually do not exist. In other words, different initial states basically lead to the same final state if the system is given enough time to evolve. Such a difference is of importance as it reveals the roles of initial state and mutation in short-term and long-term system evolution respectively. We shall present evidence to support our argument and have some brief discussions on the possible reasons leading to such different conclusions, as well as the implication of our observations in real-life applications.

The rest part of this report is organized as follows. The Deffuant model and its main extensions, namely link rewiring and mutation, are defined in Section II. Simulation results and discussions are presented in Section III. Finally, Section IV concludes the paper.

## II. Deffaunt Model without Noises

Deffuant model assumes that opinions are continuously distributed within the interval [0, 1]. At each time step $t$, a node $A$ is randomly selected together with its random neighbor $B$. Denote their opinions as $o(t, A)$ and $o(t, B)$, respectively. If the difference between these two opinions is less than a given tolerance $d$, they make consensus according to the following rules:

$$\begin{cases} o(t+1, A) = o(t, A) - \mu[o(t, A) - o(t, B)]; \\ o(t+1, B) = o(t, B) + \mu[o(t, A) - o(t, B)]. \end{cases} \quad (1)$$

A smaller value of $\mu$ may slow down the evolution process while different values of $\mu$, as long as it is within the range of (0, 1/2), is believed to lead to the same final steady state [8]. Hereafter, we use $\mu = 1/2$ as that in most of the existing works.

Noise/mutation was firstly introduced into Deffuant model in [12]. Specifically, in each time step $t$, a randomly selected node has a probability $p$ to mutate and adopt another randomly chosen opinion, following a certain basal distribution of preferred opinions in such random opinion changes.

## III. Simulation Results and Discussions

Fig. 1 [16] illustrates the typical steady state of opinion evolution where there exist consensus making, link rewiring and opinion mutation in the system. The initial state of the system has a uniform opinion distribution, and the basal distribution of preferred opinions in opinion mutation is also uniform. Let $p$ denote the probability that a node shall has opinion mutation, and $w$ the probability of link rewiring. Fig. 1 illustrates a few different cases with different values of $p$, $w$ and $d$. It can be seen that the system evolves into a few communities holding similar opinions with a bell-curve style opinion distribution centering around a certain "community consensus"; the opinion mutation, meanwhile, keeps the different communities connected. This is importance as such connections keep the shifting of nodes between different opinion communities feasible; otherwise the system may be split into a few disconnected communities, each of which holding its single-value community consensus, as that has been illustrated in quite a few studies (e.g., [10]).

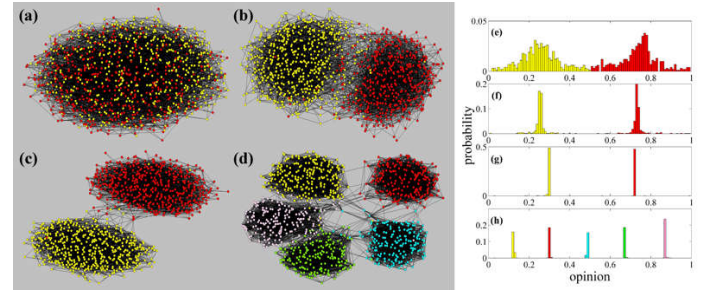

Fig. 1 [16]. Network structures at the steady state where (a) $d = 0.25$, $p = 0.1$, $w = 0.5$; (b) $d = 0.25$, $p = 0.01$, $w = 0.5$; (c) $d = 0.25$, $p = 0.001$, $w = 0.5$; and (d) $d = 0.1$, $p = 0.001$, $w = 0.5$; and their corresponding opinion distribution where (e) $d = 0.25$, $p = 0.1$, $w = 0.5$; (f) $d = 0.25$, $p = 0.01$, $w = 0.5$; (g) $d = 0.25$, $p = 0.001$, $w = 0.5$; and (h) $d = 0.1$, $p = 0.001$, $w = 0.5$. The network starts as an ER random network with a size of $N = 10^3$ and an average nodal degree of $\langle k \rangle = 10$. For Figs. (a), (b) and (c), nodes in yellow and red respectively hold opinions within range of [0, 1/2] and (1/2, 1]. For Fig. (d), nodes in yellow, red, blue, green and pink respectively hold opinions within the range of [0, 1/5], (1/5, 2/5], (2/5, 3/5], (3/5, 4/5] and (4/5, 1].

We now move forward to illustrate on the effects of initial state on final steady state of the system. We present two example cases:

**Example 1:** Erdos-Renyi (ER) random network with a network size of $n = 10,000$ and an average nodal degree of $\langle k \rangle = 20$. We let tolerance threshold $d = 0.25$, rewiring probability $w = 0.5$ and mutation probability $p = 0.1$. Snapshots are presented for $t=0$, $t=50,000$, $t=150,000$ and $t=1,500,000$ respectively. Figs. 2 and 3 show system evolution where the initial state of opinion has a uniform and a power-law distribution respectively. It can be observed that the when the evolution time is long enough, the two cases with different initial states finally converge to virtually the same final state.

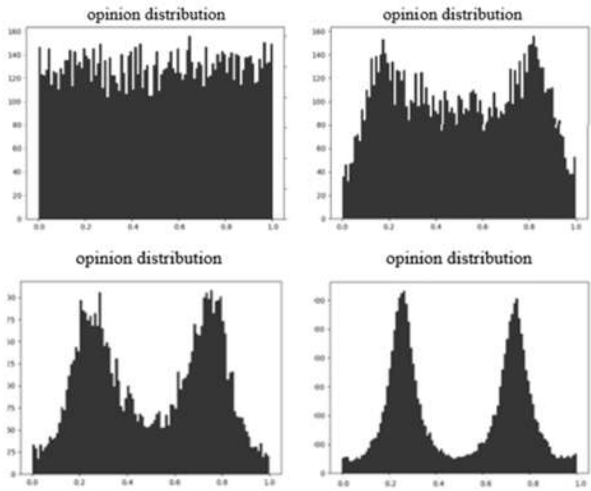

Fig. 2. Opinion evolution over time; the initial state has a uniform opinion distribution.

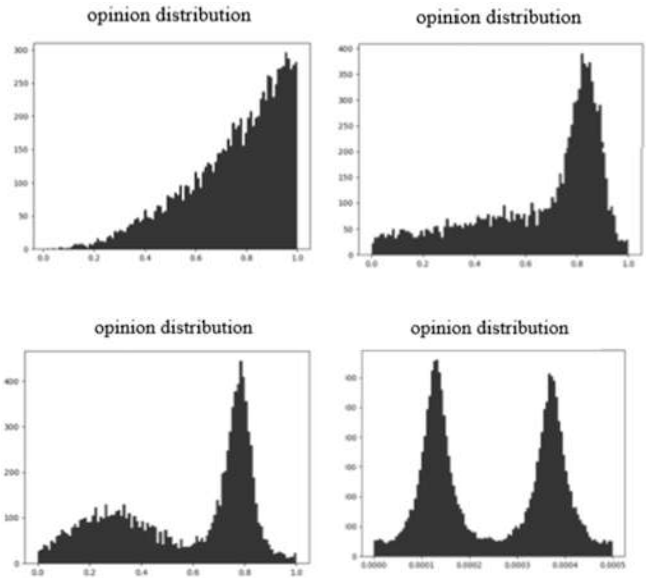

Fig. 3. Opinion evolution over time; the initial state has a power-law distribution of opinion where the exponent $\gamma = 3$.

In our extensive numerical experiments, even when the mutation probability is rather low and the rewiring probability is relatively much higher (in which case we may expect that the communities are largely isolated from each other with only sparse connections in between), systems with different initial states may still evolve into the same final state, only that the evolution time may become extremely long. An example is presented below.

**Example 2:** The tolerance threshold $d = 0.1$ and the mutation probability $p = 0.001$. All the other parameters remain the same as those in Example 1. The initial state has a power-law distribution where $\gamma = 3$. The results are presented in Fig. 4. As we can see, even when the mutation rate is rather low, after sufficiently long time (when $t = 30$ millions), system evolution still converges to the same final state as that of the case where the initial distribution has a uniform distribution.

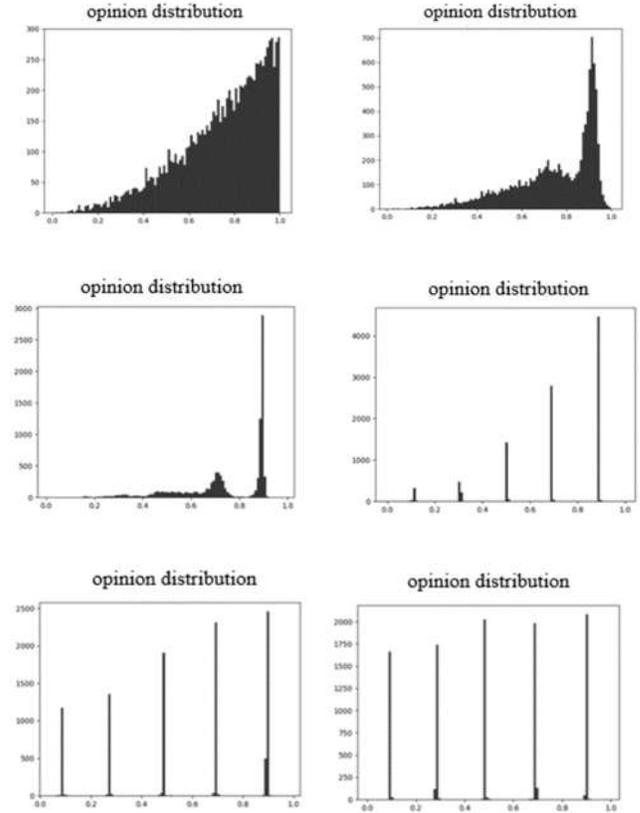

Fig. 4. Opinion evolution over time when $t$ = 0, 100k, 500k, 2mil, 10mil and 30mil, respectively. The initial state has a power-law distribution of opinion where $\gamma = 3$.

We have tested a few other cases and the conclusion remains the same that the initial state has virtually no effects on the final state of the system. Some discussions are therefore needed to explain why our observation is different from the conclusion proposed in [13]. We have good reasons to suspect that the simulations in [13]] were terminated too early (e.g., when $t = 50,000$ as stated on pp. 139 of [13]), long before the system could converge, especially when with a low mutation rate. A rule of thumb to decide the number of steps for the system to evolve into final state is to set it a few times as big as $N/p$, where $N$ denotes the number of network nodes, and $p$ the mutation probability. In this way, each node has a fair chance to have random mutation at least once. Actually in systems with consensus making and link rewiring, it takes each node to have an average of a few times of mutation before the system state converges. When the simulation time is set to be long enough, initial state shall have virtually no observable effects on the final state of the system.

Meanwhile, it shall be emphasized that the another important statement in [13] remains to be valid: it is the basal distribution of the preferred opinions that plays a critical role in deciding the steady-state opinion distribution. An illustrative example is presented in Fig. 5, where we present final steady state of two systems with (a) a power-law initial distribution and a uniform basal distribution of preferred opinions; and (b) a uniform initial distribution and a power-law basal distribution of preferred opinions, respectively. It can be clearly observed that it is the basal distribution that largely decides the final steady state of opinion distribution.

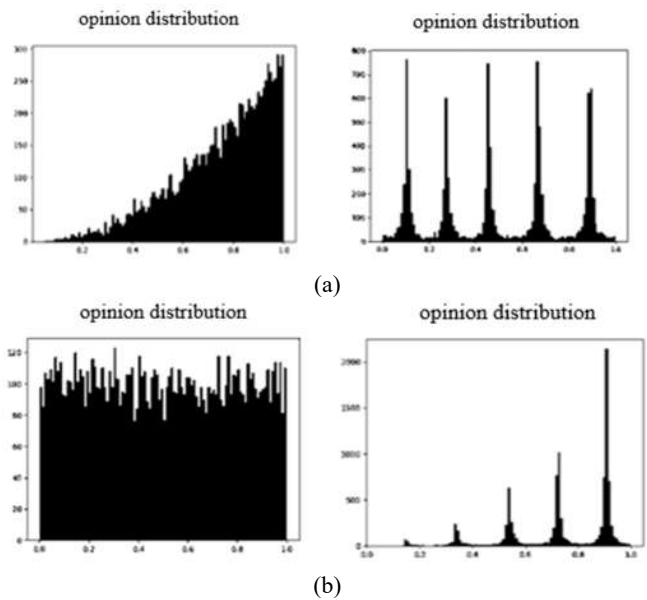

Fig. 5. Initial and final distributions of opinions in systems with (a) a power-law initial distribution and uniform basal distribution; and (b) a uniform initial distribution and power-law basal distribution

Note that though our extensive numerical simulations have shown that the initial state has virtually no effects on final steady state of opinion distribution in complex social networks, such a conclusion holds only when the evolution time is long enough. In many cases, especially in cases when the mutation probability is very low, the evolution could easily take millions of steps to converge. In real-life applications, however, a complex social system may seldom, if not never, get a chance to evolve for such a long time as a standalone system before new factors and/or outside impacts come in, driving the opinion evolution to a slightly or drastically different evolution path. This explains why in the real life, our intuitive feeling is that different initial states typically indeed make differences to the final state of system: real-life systems seldom have a chance to achieve the "final" steady state. In systems with stronger noises, however, the effects of the initial state may become level significant as such systems converge to the steady state more quickly. This, again, may match our intuitive feeling that a more "chaotic" system's evolution is less relevant to its original state some time ago.

## IV. CONCLUSIONS

In this paper, we studied on a small but nontrivial issue in opinion formation and evolution in complex social networks, namely the effects of initial state on the final steady state of opinion distribution. We argued that the initial state has virtually no effects on the system's final steady state. A few examples were presented as evidence to support our arguments, and possible reasons leading to the long-term belief that initial state does affect final steady state were briefly discussed. We pointed out that since real-life social systems seldom get a chance to converge to a steady state, effects of initial states may not be neglected altogether in real-life applications. We also emphasized on the important role that the basal distribution of preferred opinions plays in deciding the short-term and long-term evolution of complex social opinion systems.

As real-life systems are almost always *evolving* rather than *converging* to a steady state, finding a set of metrics that could conveniently reveal the transient state during the system evolution will be of our future research interest.

ACKNOWLEDGMENT

This study is partially supported by Ministry of Education, Singapore, under contract MOE2016-T2-1-119.

REFERENCES

[1] A. Pluchino, V. Latora, and A. Rapisarda, "Compromise and synchronization in opinion dynamics," *Eur. Phys. J. B*, vol. 50, no. 1–2, pp. 169–176, Mar. 2006.
[2] K. Sznajd-Weron and J. Sznajd, "Opinion evolution in closed community," *Int. J. Mod. Phys. C*, vol. 11, no. 6, pp. 1157–1165, 2000.
[3] C. Castellano, S. Fortunato, and V. Loreto, "Statistical physics of social dynamics," *Rev. Mod. Phys.*, vol. 81, no. 2, pp. 591–646, May 2009.
[4] Noah E. Friedkin, Anton V. Proskurnikov, Roberto Tempo, and Sergey E. Parsegov, "Network science on belief system dynamics under logic constraints," *Phys. Rev. Lett. Phys. Rev. B Phys. Rev. Lett. Nat. Phys. Nat. Phys. Phys. Rev. Lett. Phys. Rev. Lett*, vol. 112, no. 109, pp. 41301–256802, 1994.
[5] C. Castellano, D. Vilone, and A. Vespignani, "Incomplete ordering of the voter model on small-world networks," *EPL (Europhysics Lett.*, vol. 63, no. 1, p. 153, 2003.
[6] V. Sood and S. Redner, "Voter model on heterogeneous graphs," *Phys. Rev. Lett.*, vol. 94, no. 17, p. 178701, 2005.
[7] S. Galam, "Minority opinion spreading in random geometry," *Eur. Phys. J. B-Condensed Matter Complex Syst.*, vol. 25, no. 4, pp. 403–406, 2002.
[8] G. Deffuant, D. Neau, and F. Amblard, "Mixing beliefs among interacting agents," *Adv. Complex*, 2000.
[9] G. Deffuant, "Comparing Extremism Propagation Patterns in Continuous Opinion Models," *J. Artif. Soc. Soc. Simul.*, vol. 9, no. 3, p. 8, 2006.
[10] B. Kozma and A. Barrat, "Consensus formation on adaptive networks," *Phys. Rev. E - Stat. Nonlinear, Soft Matter Phys.*, vol. 77, no. 1, 2008.
[11] J.-D. Mathias, S. Huet, and G. Deffuant, "Bounded Confidence Model with Fixed Uncertainties and Extremists: The Opinions Can Keep Fluctuating Indefinitely," *J. Artif. Soc. Soc. Simul.*, vol. 19, no. 1, p. 6, 2016.
[12] M. Pineda, R. Toral, and E. Hernandez-Garcia, "Noisy continuous-opinion dynamics," *J. Stat. Mech*. (2009), p08001.
[13] A. Carro, R. Toral, and M. S. Miguel, "The role of noise and initial conditions in the asymptotic solution of a bounded confidence, continuous-opinion model," *J. Stat. Phys.*, vol. 151, no. 1-2, pp. 131-149, Apr. 2013.
[14] Y Yu and G. Xiao, "Influence of Random Opinion Change in Complex Networks," *Proc. IEEE DSP'2015*, July 2015
[15] Y. Yu and G. Xiao, "Preliminary study on bit-string modeling of opinion formation in complex networks, " *Proc. HUSO'2015*, Oct. 2015
[16] Y. Yu, G. Xiao, G. Li, W. P. Tay, and H. F. Teoh, "Opinion Diversity and Community Formation in Adaptive Networks," *Chaos*, vol. 27, no. 10, 10315, Oct. 2017.